\documentclass[11pt]{article}
                \usepackage{epsfig}
                \usepackage{amsmath}
                \usepackage[square]{natbib}

\begin{document}

              \title{Collaborative Virtual Queue:\\
              Fair Management of Congested Departure Operations and Benefit Analysis.}

                \author{Pierrick Burgain \thanks{Corresponding author, School of Electrical and Computer
                Engineering,
              Georgia Institute of Technology, Atlanta, GA, pburgain@gatech.edu},
              Eric Feron \thanks{School of Aerospace Engineering, Georgia Institute of Technology, Atlanta, GA, feron@gatech.edu},
              John-Paul Clarke \thanks{School of Aerospace Engineering, Georgia Institute of Technology, Atlanta, GA, johnpaul@gatech.edu} \\
  }

\date{\today}
\maketitle

 \begin{abstract}

                Due to the stochastic nature of departure operations, working at full
                capacity makes major US airports very sensitive to
                uncertainties. Consequently, airport ground operations face critically congested
                taxiways and long runway queues.

                In this report, we show how improved management of departure operations from the ready-to-push-back time to
                the wheels-off time can potentially yield significant benefits to airlines and air traffic services. We develop
                a Collaborative Virtual
                Queue to enable better optimization capabilities during congested situations while taking
                into account the laissez-faire competitive
                environment.
                Results are evaluated using a departure system
                model, validated using current statistics and
                previous studies.
                First, the Collaborative Virtual Queue enables keeping aircraft away from runway queues, which increases
                wheels-off time predictability.
                Second, holding aircraft enables last-minute intra-airline flight switching. This creates new optimization
                 capabilities for airlines i.e. it gives airlines the flexibility to prioritize their flight sequence in real-time.
                These capabilities
                 are illustrated by the trade-off between minimizing the average
                passenger waiting time and minimizing the level of unfairness between aircraft of the same airline.
                For instance, airlines could choose to decrease by up to 15\% their average
                passenger waiting time by prioritizing heavy planes over small planes when the taxiway system is congested.

\end{abstract}

\clearpage

\tableofcontents

\clearpage

 \listoffigures

\clearpage

        \section{Introduction}

The U.S. National Airspace System (NAS) is expected to accommodate
  between 2 and 3 times today's traffic level by
 2025. Along with a physical increase of airport capacity, the
 future
generation of air transportation system will require new concepts of
operations to handle the growth. Even though the development of
smaller regional airports is expected, major airports will always
have to run at full capacity. Some of them, in the NAS, will not be
able to expand their capacity enough to sustain the increasing
demand. Such airports are bound to become and remain air traffic
bottlenecks. Airports such as New-York Laguardia will be physically
restrained by the lack of space and the impossibility to add new
runways or ramps.

To tackle this structural and organizational issue, the NextGen
concept of operations \citep{nextGen} encourages research on surface
traffic operations aimed at lowering emissions and improving surface
traffic planning (table \ref{tab:NextgenResearch2}). At the same
time, the NextGen report also favors reducing the need for
government intervention and control of resources (\citep{nextGen},
lines 1709 to 1726).

\begin{table}[h]
\begin{center}
    \begin{tabular}{|p{0.1\linewidth}|p{0.4\linewidth}|p{0.4\linewidth}|}
    \hline
    \textbf{Ref} & \textbf{Line Reference}& \textbf{Issue}\\
    \hline
     \multicolumn{1}{|c|}{R-46} & At times of peak demand, major airports conduct Super-Density
Operations in which capacity-enhancing arrival and surface
procedures are implemented to maximize runway throughput.
     & How will we design and implement systems
to be resilient to failures and robust to
operator error?\\
    \hline
\multicolumn{1}{|c|}{R-135} & Advanced capability to integrate and
balance noise, emissions, fuel burn, land use, efficiency, and costs
effects of alternative measures and alternatives allow selection of
optimum operational modes, mitigation strategies, and surface
planning procedures. & Identify improvements to surface planning
procedures
to reduce emissions.\\

\hline
  \end{tabular}\caption{NextGen research and policies issues}\label{tab:NextgenResearch2}
\end{center}
\end{table}

In the mean time, in Europe, EUROCONTROL is developing a Departure
Manager (DMAN) system \citep{dman}. The DMAN is a centralized
concept of operations for surface traffic management. Though the
DMAN would hardly fit the US decentralized and laissez-faire
environment, its development introduces the use of Collaborative
Decision-Making as a tool to manage departure operations. According
to \citep{dman} ,the DMAN ``keeps the number of aircraft on the
taxiway at an optimal level'' and ``keeps the taxiways open for
other traffic without blocking stands for arrivals, reduces
controller workload, improves punctuality and predictability,
facilitates co-operation between aerodrome ATC, airlines and airport
operators, enhances CFMU [Central Flow Management Unit
slot-revisions] and slot-compliance, and exploits the departure
capacity of the respective runway''.

Most studies on airport operations agree on the importance of
controlling gate-holding and push-back sequencing. Optimization
methods to increase airport throughput and reduce surface
inefficiencies were developed. Anagnostakis, Clarke, Bohme, and
Volckers \citep{oldVq} focused on sequencing and scheduling, Capozzi
\citep{AUTOMATED-AIRPORT-SURFACE-TRAFFIC-CONTROL} on surface
automated traffic control, and Balakrishnan \citep{balaq} on
coordinated surface operations from the gate to the take-off.

Other research accurately described the departure operations as
queueing processes. Feron, Delcaire, and Pujet \citep{log99} studied
a queueing input-output model,
 Clarke, Feron, Carr and Evans \citep{modelingQ} worked on modeling
and controlling queueing dynamic under severe flow restrictions, and
Husni, Clarke, Bhuva and Kang \citep{taxioutTimeEst} developed a
queueing model for taxi-out time estimations. These studies focused
on the stochastic nature of ground operations, and pointed out the
difficulty of a fine tuning of surface operations in congested
situations. Furthermore, optimizing departure operations, especially
in congested situations, is complicated by the competitive nature of
departure operations. First, airlines have information privacy
policies that make collaboration delicate. Second, airlines want to
keep as much control as they can in departure operations.

This paper studies and estimates the potential benefits of using a
Collaborative Decision Making concept which can preserve the
decentralized laissez-faire environment of U.S airports: The
Collaborative Virtual Queue (CVQ). The CVQ uses virtual queueing to
hold planes away from runway queues and enable last-minute plane
switching . We measure the creation of optimization capabilities and
the improvement of take-off prediction robustness.

This paper is organized as follow: First the congestion problem is
described. Second, Collaborative Virtual Queue operations are
defined, explained, and compared to the actual Ground Delay
Programs.
 Then, the Boston Logan airport departure operations are modeled,
and calibrated using current flight data as well as taxiway maps.
The last part focuses on congested situations and quantifies the
optimization capabilities created by the Collaborative Virtual Queue
(CVQ).

\section{Problem Description}

In major airports, airlines and air traffic service providers
strategically optimize their operations to operate at full capacity
during activity peaks. To do so, they assume good weather
conditions. In other words, when the departure and arrival schedules
are created, it is assumed that the available arrival and departure
rates will not be limited for any reason other than regular runway
capacity and safety constraints. Moreover, airlines sometime
over-schedule flights regardless of airport resources. In addition
to tight planning, the queueing nature of surface operations implies
that there are enough planes out to ensure a constant takeoff demand
reservoir \citep{modelingQ} at runway thresholds. Consequently major
airports are very sensitive to unexpected events that disrupt the
throughput capacity. Congestion happens as a natural combination of
tight operation planning and unpredictable events (for instance bad
weather resulting in a first fix closure , i.e. closure of a
departure route in the Terminal Radar Approach CONtrol (TRACON)
area). At busy airports such as the Atlanta Airport, it is not
surprising to have queues of 30 or 45 aircraft waiting to take off
at runway thresholds.

To deal with congestion, the FAA, airports, and airlines have
several options.

\begin{enumerate}

\item They can try to eliminate congestion by:
\begin{enumerate}

\item Physically increasing their capacity (by building more runways, or by using microwave
radar technology to monitor wake-vortex and reduce
 aircraft separation restrictions).

\item Forcing restrictions on departure planning, using slot restrictions.

\end{enumerate}

\item They can also optimize operations by:
\begin{enumerate}

\item Centralizing operations around airport authorities to ease collaboration, optimize
throughput, and lower inefficiencies.

\item Using collaborative decision making to improve operations while respecting the competitive environment.

\end{enumerate}

\end{enumerate}

Solution 1-a, increasing airport capacity, is not always applicable.
For example, at New York LaGuardia, there is not enough space to add
any runway. Even when applicable, the demand for traffic stays high
and congestion is reduced, but not suppressed. Solution  1-b, using
slots restrictions, is enough to avoid obvious schedule overloading
\citep{SizeMatterAtLaguardia}. This method is necessary in some
cases, and there are currently 4 federally slot-controlled airports
(New York LaGuardia, New York JFK, Reagan Washington National, and
Chicago O'Hare). It prevents highly congested situations that could
impact the whole National Airspace System. However slot restricted
airports assume a constant departure rate which can only be achieved
in good weather conditions. In other words, slot restrictions do not
avoid unplanned throughput disruptions triggered by events such as
frequent changes in wind direction (forcing the airport control
tower to switch configuration several times during a short period),
Ground Delay Programs, or simply first fix closures due to bad
weather. Point 2-a, centralizing operations to ease collaboration,
gives much power to airport authorities and air traffic services
providers. This solution is far from usual operations in U.S.
airports which have highly decentralized operations. Indeed private
airlines control their ramp operations until aircraft are ready to
push back. The control tower gives airlines clearance to push back
on a First Come First Serve basis, as soon as their aircraft can
physically push back and exit the ramp. Consequently, attempts to
take control of push-back sequences and ground resources face high
resistance from airlines. Besides, centralizing operations reduces
the set of private airline responsibilities and suppresses the
laissez-faire environment. Thus, it reduces free market incentives
that continuously help improve operations and adapt services to
passengers' needs.

Collaborative Decision Making (CDM), Point 2-b, can help airport to
reach an intermediate solution, that will not centralize operations.
This paper studies the potential benefits of CDM applied to ground
operations during congested situations. CDM has already proven its
efficiency in Flow Management with Flight Schedule Monitor (FSM).
Indeed, FSM is a system whose main role is to implement Ground Delay
Programs when, at an airport, the arrival demand exceeds the arrival
capacity. FSM enables airlines to collaboratively reschedule their
flights to adapt the arrival demand to the reduced arrival capacity
at the airport, as described in further details in section
\ref{GDP}.

 The competitive environment plays an important role. In most airports,
 airlines compete to both meet their on-time
performance as well as to use airport resources. Therefore, in the
absence of regulating policy, airlines react to a limited throughput
capacity by trying to push back earlier \citep{minDepUncert} to get
their planes in the queue as early as possible. Once aircrafts are
queueing on the taxiway system, airlines lose optimization
capabilities. Indeed, by ``racing for the runway'', they lose the
option to switch their flights and reorganize their takeoff
sequence.

To manage congestion, any decentralized operation optimization
scheme first needs to find a solution that provides airlines
motivation for not stacking their planes in a saturated runway
queue. This could be achieved by motivating airlines to keep planes
at the gate or use additional parking spots to park planes while the
departure system is saturated. Observations of a freight company
operating as the sole operator at night at the Memphis airport
confirmed that holding planes is the corner stone of optimizing
congested departure operations. Indeed, at night at the Memphis
airport, ramp operations are dominated by one large freight company
which can optimize the number of planes taxiing and queueing for
each runway. Holding aircraft, along with the control of runway
queue lengths, enables the freight company to switch flights with
respect to their priority level. This real-time optimization is
especially useful when the departure rate is disrupted because it
allows the most valuable aircraft to take off with a lower delay.

The biggest challenge faced by holding planes from queuing on the
taxiway is space availability requirements for new arrivals (at the
gate or on the taxiway). However, in most airports, two phenomena
guarantee that free parking capacity could be exploited. First, gate
occupation and parking space requirements are cyclical, everyday
there is a time when the arrival rate is lower than the departure
rate, this creates temporary capacity which could be used to hold
aircraft. Second, airports capacity grows cyclically, when
additional ramps are built, or taxiway is extended. The sudden
increase in parking capacity allows more holding capacities,
especially as the maximum departure rate remains unchanged until a
new runway is built. Congestion will always happen, however CDM
concepts can help increase flexibility of the departure flight
sequence, yielding significant optimization capabilities.

\section{The Collaborative Virtual Queue}

\begin{figure}[ht]
\centering
\includegraphics[scale = .55]{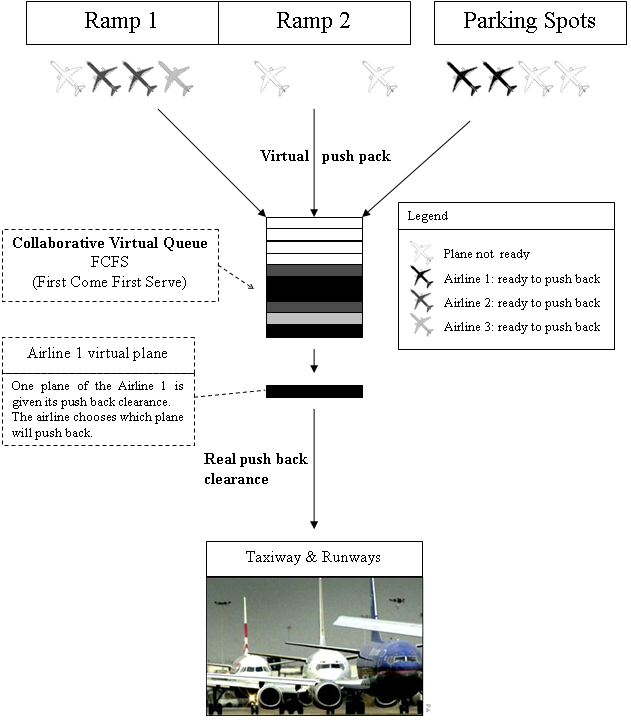}
\caption[Collaborative Virtual Queue]{Collaborative Virtual Queue}
\label{fig1}
\end{figure}

The Collaborative Virtual Queue (CVQ) is a queuing mechanism whereby
pushback requests are stacked in a virtual queue, and planes are
hold from taxiing to runway queues. Planes are hold at the gate or
at tarmac parking spots. The queue is continuously skewed by
pushback clearances from the Air Traffic Service Providers who
decide if the departure system can handle additional traffic without
sacrificing safety or performance. Figure \ref{fig1} illustrates the
CVQ concept.

 The Collaborative Virtual Queue aims at enabling airlines to
 hold planes away from taxiway and runway saturated queues.
  By doing so, the CVQ would create
last-minute flight switching capabilities, that airlines could
exploit to improve their real-time departure operations. Besides,
this mechanism would increase accuracy of taxiing time predictions
as it would reduce the number of planes queueing on the taxiway.

As a result, when ground operations are competitive and congested, a
CVQ would help control the level of taxiway congestion, while
distributing to airlines their fair share of departure capacity and
flexibility.

\clearpage

\subsection{Time frame}

The Collaborative Virtual Queue manages planes at the ramp between
the ready-to-push-back time and the actual push-back time.

Uncertainties in turn times are an issue for any departure
optimization process that would need to forecast precisely the plane
ready time. Carr and Theis \citep{pbForecast} described the limit of
push-back forecasting as the limit of turn times predictability for
future operations. They conclude that future Air Traffic Management
processes must be designed with robust mechanisms for coping with
push-back time variance. Supporting their study, Satish et al.
\citep{minDepUncert} also observed that these uncertainties in ready
to push-back times and in taxi-out times motivate airlines to use a
time buffer in their push-back forecast when they file their flight
plan.

As a result, to optimize congested operations while equitably
respecting the competition, planes will participate in the
Collaborative Virtual Queue only once there is no doubt left on
their ready time, e.g., when the pilot asks for push-back clearance
or when the doors are closed.

\subsection{Sequence of operations} \label{seqOfop}

The section describes the CVQ concept of operations in details
(Figure \ref{fig1}); it follows the sequence of events undergone by
an aircraft.

\begin{itemize}

\item When a plane on the ramp asks for a push-back clearance, the plane either stays at the
gate or moves to an alternative parking spot. A matching virtual
plane is created. This is a virtual push-back.

\begin{description}

\item{\textbf{Virtual Plane}}

A Virtual Plane (VP) is a floating push-back clearance slot. It is a
piece of information that identifies the airline which just got a
plane ready. The VP contains the virtual push-back time. It can
include any other aircraft information regarding the optimization of
push-back sequencing, this information is private to the airline.

\end{description}

\item The virtual queue administrator stacks the virtual plane in
the CVQ. There is one CVQ for all runways. Indeed, for the purpose
of this study, it is assumed that the Air Traffic Control Tower
balances runways by changing first fix runway allocations. As a
result the plane pushes back toward the runway with the least number
of planes (already queueing and/or taxiing toward it). Further study
will integrate unbalanced runways managed by multiple CVQs.

\item Once the number of planes taxiing out falls below an optimal
saturation threshold, the oldest virtual plane inside the virtual
queue will be transformed into a push-back clearance for the airline
that owns this virtual plane. The airline chooses to give the real
push-back clearance to any of its planes which have already entered
the Collaborative Virtual Queue and corresponds to the virtual plane
description. This respects the competitive environment currently
generated by the First-Come-First-Serve (FCFS) policy applied by Air
Traffic Control Towers in most U.S. airports. Notice that this
optimization capability does not interfere with other departure
optimization processes happening before the ready time or after the
push-back. It creates a buffer providing flexibility for the airline
to switch flights with different degrees of priority in their
take-off sequence. Figure \ref{fig1} illustrates the CVQ based on a
simple First-Come-First-Served policy.

\item If the departure system has enough capacity to receive all planes
asking for push-back without saturating the system, the CVQ will be
transparent to the departure operations.

\end{itemize}

\clearpage

\subsection{Comparison with Ground Delay Program} \label{GDP}

The CVQ has many common points with the Ground Delay Program (GDP).
The GDP is an initiative to manage the traffic flow into an airport.
More precisely, it is a tool developed to decrease the rate of
incoming flights into an airport as soon as it becomes highly
probable that arrival demand will exceed capacity \citep{FSMmanual}.
This happens when the Airport Acceptance Rate (AAR) is unexpectedly
low due to weather or other conditions. For instance the Severe
Weather Avoidance Procedures (SWAP) relies on the GDPs as a method
available to hold flights on the ground.

 Collaborative Decision Making within the GDP is enabled by the Flight
 Schedule Monitor (FSM).
 This collaborative system \citep{FSMmanual} is used by airlines, airports, and the Air Traffic Control System
Command Center (ATCSCC) to implement Ground Stop, Ground Delay
Program, and Airspace Flow Program strategies. Airports and airlines
share their schedule information to enable common situational
awareness.

FSM looks at all the known demand (scheduled flights and flight
plans that have been entered in the Enhanced Traffic Management
System (ETMS)). The core algorithm tries to equitably assign delays
to all known flights in order to match the arrival demand to the
airport arrival capacity. Airlines can watch arrival demand levels
and ask for specific arrival slots but they do not have information
on their competitors.

Although GDP operates on a strategic level (1 to 6 hours before
arrival) and the CVQ on a tactical-opportunistic level (0 to 20
minutes before push-back), they have many similarities. The GDP
creates departure slots to accommodate limited arrival capacity
while the CVQ creates push-back slots to accommodate limited
departure capacity. Both programs are aimed at optimizing demand
with respect to a limited capacity in a competitive environment, and
both reorganize operations using virtual queues (arrival slots in
FSM form a virtual queue).

\clearpage

\subsection{Data source and setting of parameters}\label{dataSource}

Boston Logan Airport departure operations were first modeled without
any Collaborative Virtual Queue, using the following sources: the
existing description from the studies \citep{log99, modelingQ,
taxioutTimeEst}, a study on push-back forecasting
\citep{pbForecast}, current data from Airline Service Quality
Performance (ASQP), Enhanced Traffic Management System Counts
(ETMSC), Aviation System Performance Metrics (ASPM), and Google
Earth.

The ASPM database provided us with flights' historic times with a 60
second resolution for all flights which departed from the Boston
Logan airport. These times include: gate out, taxi-out time, and
wheels off times, allowing us to calibrate the departure rate,
taxi-out speed, and push-back rate.

The flight carrier code given by the ASPM data enabled us to
identify the participation of every airline in departure operations,
from January 1st to September 30th 2006, at the Boston Airport. For
privacy purposes, airline ICAO codes were replaced by grades from AA
to CP, AA being the airline which had the most flights.

Table \ref{tab:ICAO} shows the top 10 airlines (out of 67) and gives
their ranking codes. Their cumulative participation represents 74\%
of the total departure operations during this period.

\begin{table}[h]
\begin{center}
\begin{tabular}{|c|l|c|c|}
 \hline
Airline Grade & Nationality & Percentage of Departures\\
 \hline
 AA &  USA & 10.60\% \\
AB &  USA & 9.27\% \\
 AC &  USA & 9.04\% \\
 AD &  USA & 8.95\% \\
 AE &  USA & 8.14\% \\
 AF &  USA & 6.77\% \\
 AG &  USA & 5.93\% \\
 AH &  USA & 5.87\% \\
AI &  USA & 5.87\% \\
 AJ &  USA & 3.73\% \\

 \hline
 \end{tabular}\caption{Airlines departure mix}\label{tab:ICAO}
\end{center}
\end{table}

\begin{figure}[ht]
\centering
\includegraphics[scale = .6]{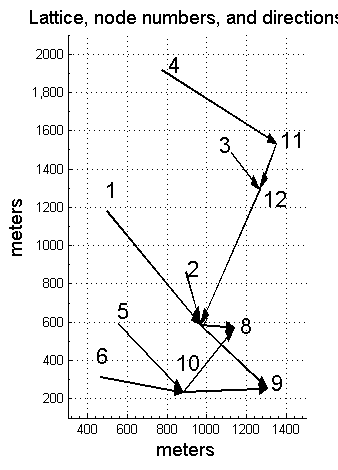}
\caption[Lattice - Taxiway topology]{Lattice - Taxiway topology,
departures on runways 9 (node 9)} \label{taxiwayLattice}
\end{figure}

 The taxiway topology is derived using Google Earth
meter measurements. Figure \ref{taxiwayLattice} represents the
lattice.

The number of seats per aircraft at Boston is derived from the
database Enhanced Traffic Management System Counts (ETMSC, available
at http://www.apo.data.faa.gov) combined to the ETMS Equipment code
per flight, given by the ASPM database, average values are
summarized per category in table \ref{tab:fleetMix}.

\begin{table}[h]
\begin{center}
\begin{tabular}{|l|r|r|}
\hline
Plane Type & Percentage & Average Seats\\
 \hline
 Heavy & 16.73 & 214\\
 Large & 77.21 & 97\\
 Small & 6.06 & 4\\
 \hline
 \end{tabular}\caption{Aircraft departure mix}\label{tab:fleetMix}
\end{center}
\end{table}

\clearpage

\section{Modeling of Airport Departure Ground Operations}

In order to evaluate the Collaborative Virtual Queue concept of
operations proposed in this paper, we modeled the Boston Logan
airport operations, configured for departures on runways 9 and
arrivals on runway 4R. Airlines hand over control of their plane to
the Air Traffic Control Tower right after it pushed back
\citep{log99,modelingQ}. Congestion is expected on taxiways at times
of high departure rate. This model will evaluate the flexibility
capabilities enabled by last-minute plane switching and gate
holding.

\subsection{Departure System}

\begin{figure}[ht]
\centering
\includegraphics[scale = .6]{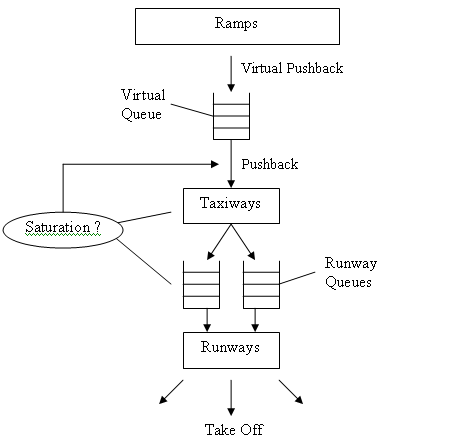}
\caption[Schematic representation of departure operations]{Schematic
representation of departure operations} \label{modelSchem}
\end{figure}

Several studies \citep{log99,modelingQ,taxioutTimeEst} demonstrated
that the degree of complexity and the number of unpredictable
factors affecting ground operations are such that ground operations
are best modeled by stochastic queueing models. This model (Figure
\ref{modelSchem}) uses the same server structure as previous
queueing models by Feron and Carr \citep{log99,modelingQ} to
generate take-offs. However, this model includes significant
improvements over previous models. First, the taxiway system is
modeled by a lattice. Aircraft taxi times are not determined
directly, instead, the model simulates aircraft motion on the
taxiway, from the gate to the runway. Second, previous models were
event-driven, this model uses a time-driven simulation with a 30
second time step. Third, this model emulates airline push-back
decisions by sweeping through different push-back policies, as
illustrated as detailed in section \ref{pushBackPolicies}.

\paragraph{Architecture}

 Figure \ref{modelSchem} illustrates the
architecture of the model. A virtual queue keeps track of ready
planes at the ramps before they push back on the taxiway. Two runway
queues link the taxiway system to each runway. The congestion level
is estimated by counting the number of planes out on the taxiway and
in the runway queues.

\paragraph{Departure Schedule}

Aircraft appear at the ramps, their push-back schedule and weight
category are determined from historic data, see section
\ref{dataSource}.

\paragraph{Model Virtual Queue}

When a plane is ready (when it appears at the ramp), a virtual plane
is created and stacked in the virtual queue (Figure
\ref{modelSchem}, according to the process described in section
\ref{seqOfop}). Once the departure system is not saturated anymore,
the company who has the oldest virtual plane is allowed to push-back
the aircraft of its choice (see section \ref{pushBackPolicies}).

\paragraph{Taxiway}

The topology of the airport taxiways, the ramp positions, and the
runway positions are modeled using a lattice. When a plane pushes
back, it takes the shortest path from the gate to the runway
threshold. They move with a stochastic speed defined as follows:
they move a constant distance at each simulation step unless an
event (generated by a bernoulli variable) stops them. This variable
was calibrate so that the distribution of taxi times would fit the
estimated real distribution of taxi times. The real taxi time
distribution was estimated by observing the taxi-out times when few
planes where taxiing out (derived from ASPM database and
\citep{log99}).

\paragraph{Runways and Takeoffs}

When aircraft reach the runway 9, they enter the departure runway
queue as described by Figure \ref{modelSchem}. The runway itself is
modeled as a server whose service rate is generated by the sum of 2
bernoulli variables. These variables are calibrated to fit the mean
and standard deviation of the take-off rate when the number of
planes out is higher than 12 (derived from ASPM database and
\citep{log99}).

\subsection{Airline Push-back Policies} \label{pushBackPolicies}

 The Collaborative Virtual Queue gives airlines the opportunity to adjust their departure
  operations by choosing what plane to push-back at the last minute.
 Their push back policy can either (i) decrease the average time their passengers waited
before take-off or (ii) minimize unfairness in waiting times among
their planes. This trade-off illustrates the optimization
capabilities generated by the Collaborative Virtual Queue.

\paragraph{}
 Two push-back policies
are at each end of this trade-off.

\begin{itemize}

\item At one end, airlines can choose not to switch any flight, i.e. apply a First-Come-First-Served
(FCFS) policy. In this case they minimize the unfairness in waiting
times between their planes.

\item  At the other end, airlines can choose to minimize the average time their
passengers spend in the planes between ready time and wheels-off
time. To do so, among the planes waiting for push-back in the CVQ,
airlines always push back their heaviest plane first. Let ``Heaviest
Plane First'' denote this policy.

\end{itemize}

\paragraph{Cost Functions}

A cost function attributes a cost to holding planes. When an airline
pushes back one of its ready planes, it chooses the plane whose
holding cost is the highest.

The FCFS policy can be implemented using any strictly monotonic
function of the holding time at the gate. One cost function which
corresponds to the FCFS policy is:

$$C_1 = TimeSinceReady$$

The Heaviest Plane First (HPF) policy can be implemented using any
cost function which respects the classification of plane sizes. One
cost function which corresponds to the Heaviest Plane First (HPF)
policy is:

$$C_2 = NumberOfPassengersInPlane$$

A tunable cost function, combining $C_1$ and $C_2$, was developed to
gradually switch from a FCFS policy to a Heaviest Plane First
policy. The parameter $\alpha$ was swept from 0 to 1. For each value
of $\alpha$, 64 days of operations were simulated using $C(\alpha)$
as the cost function for every aircraft.

$$C(\alpha) = w_1*C_1*(1-\alpha)+w_2*C_2*\alpha$$ \label{costFunct}

\paragraph{Variable definitions:}
\begin{itemize}

\item $NumberOfPassengersInPlane$: Number of passengers per plane.
It is derived from the aircraft weight category and the average
number of seats per aircraft of this category (section
\ref{dataSource}, table \ref{tab:fleetMix}).

\item $TimeSinceReady$: Time a plane has
been held at the gate since it is ready to push-back.

\item $C_1$ and $C_2$ are the costs attributed respectively to
the FCFS and HPF policies.

\item $\alpha$ is the sweeping parameter. It varies between 0 and 1. By sweeping it from 0 to 1, the
push-back policy gradually switches from the FCFS policy ($C_1$) to
the Heaviest Plane First policy ($C_2$).

\item $w_1$ and $w_2$ are respectively the weights of $C_1$ (FCFS) and $C_2$ (HPF) in the total cost
function $C$. As $\alpha$ varies, the ratio $w_1/w_2$ influences the
distribution of intermediate trade-off policies between the FCFS
policy and the HPF policy. The ratio is fixed at $w_1/w_2=4$ by
trials and errors, it yields results well spread between the $\alpha
= 0$ and $\alpha = 1$ cases.

\end{itemize}

\subsection{Set Optimal Number of Planes
Taxiing-Out}\label{setOptimalLoadLimit}

Second, the Collaborative Virtual Queue concept of operations was
implemented.

We used the method described by Pujet et al. in \citep{log99} to set
an optimal gate holding threshold:  gate holding can lower the
operating costs because costs at the gate are lower than costs
taxiing out, engines on. However the limitation of the number of
planes taxiing out (i.e., load limit) must never limit the airport
throughput.  The limit needs to be above 9 airplanes taxiing out in
order to not limit the departure rate.

\section{Results}

The the Boston Logan departure operations are simulated, configured
with departures on runway 9 and arrivals on runway 4R.

\subsection{Reduction of Taxi-out Time Uncertainty}

\begin{figure}[h]
\centering
\includegraphics[scale = .8]{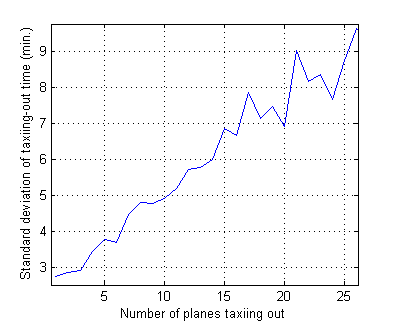}
\caption[Taxi-out time uncertainty in minutes as a function of the
number of planes already taxiing out]{Taxi-out time uncertainty in
minutes when a plane pushes back as a function of the number of
planes already taxiing out. 100 days of operations were simulated.}
\label{stdVSPlanesOut}
\end{figure}

Figure \ref{stdVSPlanesOut} illustrates the relationship between the
number of planes taxiing out and the uncertainty of the average
taxi-out time.

 When the CVQ is activated and the maximum
number of planes out set to 9, the model gives an overall taxi-out
average of 13 minutes with a standard deviation of 5 minutes, the
average number of planes out when a plane pushes back is then 7
planes out.

Similar actual operating conditions (limiting the number of planes
taxiing out) yielded similar results. Indeed, during a visit at the
Memphis airport, we conducted interviews and observed operations:
the number of planes taxiing is limited. During the day, traffic is
light. At night, one freight company controls most of the airport
ramps; as a result, the freight company limits the number of planes
taxiing out to 8 planes per runway. Therefore, the conditions at
Memphis are close to the conditions simulated for a Collaborative
Virtual Queue (CVQ) at the Boston Airport, with a limit of 9 planes
out for one runway.

To conclude, the results show that for a limit set at 9 planes out,
 the uncertainty in taxi-out time is limited to 5 minutes without limiting the airport throughput.
 Even in a competitive environment,
using a Collaborative Virtual Queue, airlines could achieve a
prediction performance close to the one achieved at Memphis airport.

\subsection{Real-time Intra-Airline Optimization Benefits}\label{results}

This section studies optimization capabilities created by the
possibility to modify the push-back sequence. Indeed, the
Collaborative Virtual Queue enables gate holding and last-minute
flight switching in the push-back sequence. The maximum number of
planes taxiing out is set to 9 in order to maximize the airport
throughput, this respects the condition given by the previous
section \ref{setOptimalLoadLimit}.

\subsubsection{Trade-off and Flexibility}

First, we illustrate the trade-off which exists between optimizing
waiting times for passengers, and minimizing unfairness among
departing aircraft. To do so, we simulate departure operations while
sweeping through different cost functions $C(\alpha)$ (section
\ref{costFunct}):

\begin{itemize}
\item $\alpha = 0$ corresponds to a FCFS push-back policy and
minimizes unfairness among departing aircraft.
\item $\alpha = 1$ corresponds to a Heaviest Plane First policy and
minimizes passenger waiting times.
\end{itemize}

For every intermediate trade-off policy, 64 days of operations were
simulated, i.e. for every $\alpha$ varying from 0 to 1 by 0.05 steps
in the cost function. We simulated departure operations assuming
departure operations were dominated by 3 independent airlines.
Indeed, at the Boston Logan Airport, 3 out of 68 airlines
represented 30\% of departure operations, from January to September
2006. In the next section we study in more details the influence of
distributions of airline departure operations on the CVQ.

On average, small planes carry  4 passengers, medium carry 97, and
heavy planes carry 214.

\begin{figure}[h]
\centering
\includegraphics[scale = 1]{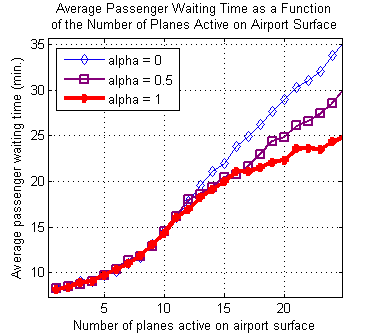}
\caption{Impact of the number of active planes on the average
passenger waiting time} \label{avWait}
\end{figure}

Figure \ref{avWait} illustrates the average passenger waiting time
\footnote{The average passenger waiting time is defined as the
weighted average of plane waiting times, using the number of
passengers per plane as the weights. The plane waiting time is the
time between ready-to-push-back and wheels-off.}, for $\alpha$ at 0,
0.5, and 1 (see section \ref{pushBackPolicies}). The top curve,
$\alpha$ = 0, shows the average passenger waiting time in the FCFS
case, i.e. when there is no optimization. The bottom curve, $\alpha$
= 1, shows the average passenger waiting time in the Heaviest Plane
First case. The curve $\alpha$ = 0.5 represents an intermediate
policy, where heavy planes have an advantage but are not
systematically more important than small planes which have been
waiting longer at the gate for push-back clearance. There is no
difference between the curves if the number of active planes
\footnote{The number of active planes is the number of planes whose
state is between ready-to-push-back and wheels-off.} is below 9,
i.e. below the gate-holding threshold, because the Collaborative
Virtual Queue has no influence on operations. However, when the
number of active planes saturates the airport departure capacity,
the Collaborative Virtual Queue gives to companies the option to use
optimization capabilities to lower the average passenger waiting
time. This is why at 20 active planes out, the curve $\alpha$ = 1 is
7.5 minutes below the FCFS $\alpha$ = 0 curve.

 Although airlines can improve the average passenger waiting time,
 there is a clear trade-off between optimizing average passenger
waiting time and minimizing unfairness between aircraft within the
same airline. This trade-off represents the new optimization
capabilities, enabled by the Collaborative Virtual Queue. Airlines
have the opportunity to use this trade-off to adjust their
operations in real-time, with respect to their business priorities.

\begin{figure}[h]
\centering
\includegraphics[scale = 1]{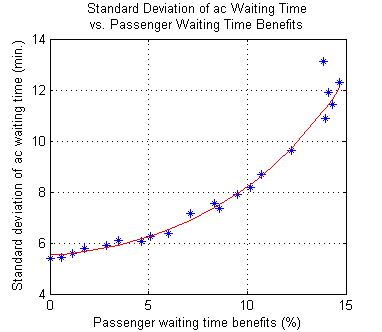}
\caption[Standard deviation of the average aircraft waiting time as
a function of the passenger waiting time benefits]{Standard
deviation of average aircraft waiting time as a function of
passenger waiting time benefits} \label{tradeOff}
\end{figure}

\begin{figure}[ht]
\centering
\includegraphics[scale = 1]{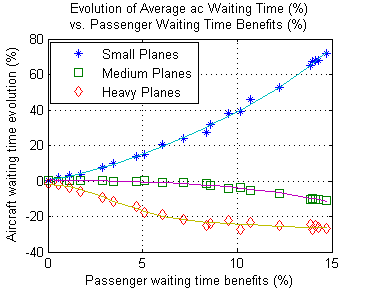}
\caption[Evolution in percent of the aircraft waiting time, for each
aircraft type, as a function of the passenger waiting time benefits
in percent]{Evolution in percent of the aircraft waiting time, for
each aircraft type, as a function of the passenger waiting time
benefits in percent. Each point represents an $\alpha$ value from 0
to 1.} \label{tradeOffPerType}
\end{figure}

\begin{figure}[ht]
\centering
\includegraphics[scale = 1]{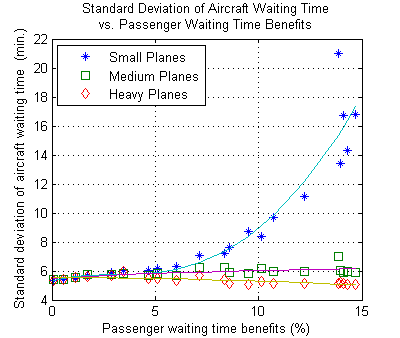}
\caption{Standard deviation of the average aircraft waiting time,
classified per type, as a function of the passenger waiting time
benefits} \label{stdPerType}
\end{figure}

Figure \ref{tradeOff}, Figure \ref{stdPerType} and Figure
\ref{tradeOffPerType} illustrate this trade-off. In the 3 figures,
plain lines represent fitted curves and points represent simulation
results. The 3 distributions where fitted using functions of the
type: $f = A*\exp{(\beta*x^2)}+B*\exp{(\delta*x)}$, where
$A,B,\beta,\delta$ are fitting parameters and $x$ is the passenger
waiting time benefits in (\%).

Figure \ref{tradeOff} illustrates unfairness between aircraft
waiting time
 as a function of benefits in passenger waiting
times . The average unfairness of waiting times between planes is
illustrated by the standard deviation of the plane waiting times, it
is minimum when there is no last-minute plane switching and
$\alpha=0$, and maximum for $\alpha=1$, for the Heaviest Plane First
policy. The percent of passenger benefits is estimated by comparing
the average passenger waiting time for a specific $\alpha$, to the
average passenger waiting time for $\alpha = 0$, in the FCFS
 case (FCFS represents
the no-optimization case and serves as a reference). Figure
\ref{tradeOff} shows that to reduce by 10\% the average waiting
time, airlines increase the standard deviation of their aircraft
waiting times by 2.5 minutes.

\paragraph{}
The next figures details how the trade-off affects every type of
plane.

Figure \ref{tradeOffPerType} displays the evolution, in percent, of
the aircraft waiting time, for each aircraft type, as a function of
the passenger waiting time benefits. The evolution rate of the
aircraft waiting time was estimated by comparing the average waiting
time of same-type aircraft to the average waiting time of all
aircraft \footnote{The expected average waiting time of all aircraft
is not affected by changes of policy. Indeed, when a company decides
to push back a plane before another one, the company departure rate
remains the same, it transfers the delay from one plane to another.
The simulation confirmed that the average plane waiting time was
constant.}. Figure \ref{tradeOffPerType} illustrates, for instance,
that a policy, yielding a 40\% increase in average waiting times for
small planes, a quasi status-quo for medium planes, and a 27\%
decrease for heavy planes, results in a 10\% decrease in the average
passenger waiting time. This policy corresponds to $\alpha=0.65$.

Figure \ref{stdPerType} is similar to figure \ref{tradeOff}, but it
details the standard deviation of plane waiting times for each type
of plane: small, medium , or heavy. One can observe that for high
values of $\alpha$,  it decreases the standard deviation of waiting
time for heavy planes, while it dramatically increases it for small
planes. For instance, a policy yielding a 10\% decrease in the
average passenger waiting time induces a slight 20 second
improvement in uncertainty of waiting time for heavy planes. In the
mean time, this policy increases the standard deviation of waiting
times by 30 seconds for medium planes, and by 3 minutes for small
planes. Considering that small planes have on average 10 times fewer
passengers than medium planes and 20 times fewer passengers than
heavy planes, this policy would represent a plausible airline
choice.

\clearpage

\subsubsection{Influence of Airline Departure Distributions}

\begin{figure}[ht]
\centering
\includegraphics[scale = .8]{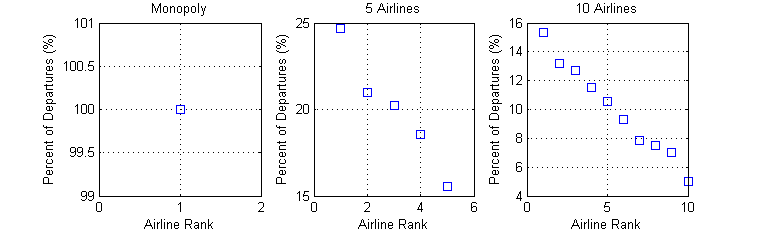}
\caption{Airlines' share of departure operations}
\label{airlinesDistri}
\end{figure}

 Airline shares of departure operations influence
optimization capabilities. To analyze their impact, results were
simulated using several distributions of airline departure
operations. Figure \ref{airlinesDistri} illustrates the 3 departure
distributions studied. The first distribution corresponds to a
monopoly, the second distribution corresponds to 5 airlines sharing
all departure resources, and the last corresponds to 10 airlines
sharing all departure resources. The monopoly was simulated by
assigning the same airline to every departure. The two other airline
distributions with 5 and 10 different airlines were obtained by
selecting the 5 or 10 first airlines (airline ranks and departure
shares are given in table \ref{tab:ICAO}) and distributing the rest
of the remaining departures to these airlines.

The case where 10 airlines share departure operations is a good
approximation of actual airline departure operations at the Logan
Airport. Indeed the first 10 airlines share 74\% of the total
departure operations at the Logan Airport, while each of the other
58 airlines which departed during the same period represent less
than 3.5\% each, totaling 26\% of departure operations.

\begin{figure}[ht]
\centering
\includegraphics[scale = .8]{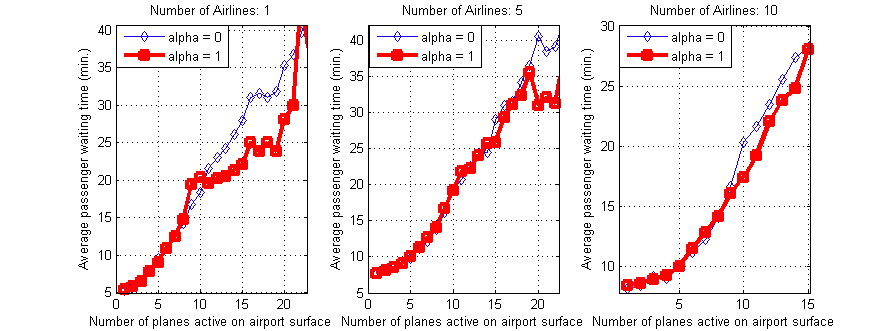}
\caption{Impact of the number of active planes on the average
passenger waiting time for 3 airline distributions}
\label{airlinesDistriCVQ}
\end{figure}

Figure \ref{airlinesDistriCVQ} shows that the average passenger
waiting time can still be significantly reduced when departure
operations are dominated by 10 airlines. Indeed, during congested
departure operations, a Heaviest Plane First policy ($\alpha = 1$)
could save up to 12\% of passenger waiting time.

 \clearpage

\subsection{Benefits Summary}

The implementation of a Collaborative Virtual Queue can lower
operating costs, and increase taxi-out time predictions. It can
generate flexibility, yielding additional optimization capabilities.
These can be used, for instance, to lower the average waiting time
per passenger, or lower the wheels-off time unpredictability for
heavy planes (containing in average 214 passengers) at the expense
of small planes (containing in average only 4 passengers).

The CVQ relates to NextGen Concepts of Operations. It provides more
flexibility to airlines and reduces the need for government
intervention. It also helps the scalability in traffic load and
demand by enabling gate holding. Thus it follows key characteristics
of the NextGen report \citep{nextGen} NextGen chap 1.2.2.1 and chap
1.2.2.9. In addition, a CVQ can potentially increase predictability
of wheels-off times by decreasing useless taxiway queueing. As such,
it helps maintaining ``efficient operations at peak capacity without
sacrificing safety'' NextGen report chap 3.1.1.

Consequently, airlines which ``participate in the collaboration
process are better able to achieve their own objectives within the
constraints imposed by overall traffic demand or short-term effects
such as weather or airspace restrictions'' NextGen report chap 2.3.

Note that the Collaborative Virtual Queue concept does not intend to
dictate a specific optimization scheme and will let airlines decide
what is best for their business model, it merely opens new degrees
of freedom.

\section{Conclusion}

In this paper, we have introduced the concept of Collaborative
Virtual Queue (CVQ): building upon previous gate holding concepts,
we explore the possibility for airlines to exploit the stack
generated by virtual queueing mechanism to reorganize their
departure sequence.

Airlines can transfer delays between different flights according to
their level of priority. The level of flexibility allowed by this
strategy has been illustrated by the trade-off between two factors,
the average waiting time per passenger, and the level of unfairness
between the airline aircraft.

The CVQ concept favors some restrictions, but only when a real
operational need exists. Because the CVQ mechanisms are based on
collaboration between private airlines, it reduces the need for
government intervention and control of resources. NextGen CONOPS
\citep{nextGen} encourage such collaborative concepts.

Further study will determine in more details the impact of different
competitive environments, of ramp topology, and the CVQ concept will
evolve to include arrival control and gate management.

\clearpage

\bibliographystyle{plainnat}
\bibliography{myBiby2}
       \end{document}